\documentclass[aps,pra,twocolumn,groupedaddress,showkeys,showpacs]{revtex4-1}
\usepackage{amsmath}
\usepackage{graphicx}
\usepackage{gensymb}
\usepackage{dcolumn}
\usepackage{graphics}
\usepackage{epstopdf}
\usepackage{amssymb}
\usepackage{bm}
\usepackage{float}
\usepackage{color}

\makeatletter
\let\cat@comma@active\@empty
\makeatother

\begin{document}
\title{Multipolar engineering of subwavelength dielectric particles \\ for scattering enhancement}


  \author{%
  S.\,D.~Krasikov$^{1}$, M.\,A.~Odit$^{1,2}$, D.\,A.~Dobrykh$^{1,3}$, I.\,M.~Yusupov$^1$, A.\,A.~Mikhailovskaya$^{1,3}$, D.\,T.~Shakirova$^1$, A.\,A.~Shcherbakov$^1$, A.\,P.~Slobozhanyuk$^1$, P.~Ginzburg$^{3,4}$, D.\,S.~Filonov$^{4}$, A.\,A.~Bogdanov$^{1,}$}
\email{a.bogdanov@metalab.ifmo.ru}



\affiliation{%
  $^1$Department of Physics and Engineering, ITMO University, St. Petersburg 197101, Russia \looseness=-1
  \\
  $^2$Electrotechnical University LETI, St. Petersburg, 197376, Russia
  \\  
  $^3$School of Electrical Engineering, Tel Aviv University, Tel Aviv 69978, Israel
  \\ 
  $^4$Center for Photonics and 2D Materials, Moscow Institute of Physics and Technology,  Dolgoprudny 141700, Russia \looseness=-1}


\begin{abstract}
Electromagnetic scattering on subwavelength structures keeps attracting attention owing to abroad range of possible applications, where this phenomenon is in use. Fundamental limits of scattering cross-section, being well understood in spherical geometries, are overlooked in cases of low-symmetry resonators. Here, we revise the notion of {\it superscattering} and link this property with symmetry groups of the scattering potential. We demonstrate pathways to spectrally overlap several eigenmodes of a resonator in a way they interfere constructively and enhance the scattering cross-section. As a particular example, we demonstrate spectral overlapping of several electric and magnetic modes in a subwavelength entirely homogeneous ceramic resonator. The optimized structures show the excess of a dipolar scattering cross-section limit for a sphere up to a factor of four. The revealed rules, which link symmetry groups with fundamental scattering limits, allow performing and assessing designs of subwavelength supperscatterers, which can find a use in label-free imaging, compact antennas, long-range radio frequency identification, and many other fields.  
\end{abstract}

%
%
\maketitle   

\section{Introduction}

Interaction of the electromagnetic waves with the matter has been a subject of intensive fundamental and applied studies over the years. Since Maxwell’s equations are proven to describe classical phenomena in a closed form, the research efforts are shifted towards more applied directions. Scattering is one of the phenomena with a far-going practical prospective, ranging from molecular spectroscopy to wireless power transfer and wireless communications~\cite{nie1997probing, lee2005dependence, jackson2004surface, schuller2009optical, hirsch2003nanoshell, song2017wireless, geffrin2012magnetic}. Thus, efficient manipulation of electromagnetic scattering underlies antenna devices, radars, and radio frequency identification (RFID) technologies. A long-standing challenge in the field remains the miniaturization of resonating elements without significant degradation of their performance. Typically, scattering cross-section (SCS) of a massive object is directly linked to its geometrical size, unless resonant phenomena are involved ~\cite{bohren1998absorption}. It might be quite contra-intuitively that SCS of subwavelength lossless structures does not depend on their sizes if the resonance condition is maintained. The latter can be achieved by either using a plasmonic resonance, increasing the refractive index of particles while reducing their sizes, or loading the antennas with additional impedances~\cite{maier2007plasmonics, krasnok2012all,mosallaei2004antenna, liu2017miniaturized, filonov2018artificial, dobrykh20204d}. These approaches are applied in the field of all-dielectric nanophotonics or GHz-range metamaterials. What does affected by resonators size is the scattering peak bandwidth, which drops significantly with the form factor - this is the celebrated Chu-Harrington limit~\cite{harrington1960effect,chu1948physical}. Since the first introduction of the concept, quite a few additional fundamental bounds (e.g. Geyi’s limit) have been derived~\cite{geyi2003method}. It becomes clear that a significant SCS enhancement in subwavelength geometries can be achieved only with spectral co-location of several resonances. This approach is directly linked to the topic of super-directive antennas, which pros and cons are comprehensively covered in a seminal Hansen’s book~\cite{hansen2006electrically}. 

Analysis of fundamental scattering bounds is well understood in application to spherically symmetric scatters~\cite{bohren1998absorption}. In this case, the far-field signatures are directly mapped on eigenmodes of a structure. The scattered field can be expanded into a series of vector spherical harmonics, forming an orthogonal basis for the vector field in three-dimensional space. Each harmonic has a characteristic far-field radiation pattern associated with the far-field of a point multipole (dipole, quadrupole, octupole, etc.)~\cite{evlyukhin2016optical}. Thus, vector spherical harmonics are often called {\it multipoles}. Due to orthogonality, each multipole represents an independent scattering channel, thus, the total SCS ($C_\text{sct}$) can be written as the sum of partial cross-sections $C_\text{sct}=\sum_{\ell,m, \sigma}C_{\ell m}^{\sigma}$. Here, $\ell$ is the orbital angular momentum and $m$ is its projection on an arbitrarily chosen z-axis and it is subject to $-\ell \leq m \leq \ell$, and $\sigma$ labels polarization. For spherically symmetric resonators, each partial  cross-section has an upper bound $C_\text{max}^\ell$ ({\it single-channel limit}), which depends only on the wavelength $\lambda$ and angular momentum $\ell$ ~\cite{foot2005atomic, tribelskii1984resonant}:
\begin{equation}
  C_\text{max}^\ell= \frac{2 \ell + 1}{2\pi}\lambda^2.
  \label{eq:limit_sphere}
\end{equation}
The scattering cross-section reaches this limit if Mie resonance condition is met. In order to surpass this single-channel limit, several Mie resonances should be spectrally overlapped.  If this condition is met, a subwavelength structure is considered as a {\it superscatterer}~\cite{ruan2010superscattering,ruan2011design}. Superscattering phenomenon is usually studied in application to high-symmetry objects, limited to spheres and two-dimensional cylinders~\cite{ruan2010superscattering,ruan2011design,qian2019experimental,mirzaei2013cloaking,raad2019multi,qian2018multifrequency,mirzaei2014superscattering}. The reason is two-fold. First, those structures have closed-form analytical solutions, which allow finding optimal (yet quite complex) conditions for scattering enhancement. Second, each resonance of a sphere corresponds to a single scattering channel as it  contributes to a single spherical harmonic. Strictly speaking, the notation of superscattering only applies to the cases when scattering channels [Eq.~\eqref{eq:limit_sphere}] can be identified and the overall cross-section can be assessed versus the limit (typically dipolar one ($\ell=1$) is the most frequently used measure).  
If the symmetry of a scatterer is broken, the mapping between multipole expansion and eigenmodes becomes more complicated. In this case, the whole concept of superscattering should be revised since the single-channel limit defined for spherical objects is not valid for non-spherical ones.
In fact, different multipoles of the incident field can be scattered into one channel, contrary to the case of high-symmetry resonators, where there is no rescattering between the channels (see Sec.~\ref{sec:single_channel_limit}). Hence, assessing scattering performances of non-symmetric structures requires developing original classification tools, paving ways to new approaches to SCS enhancement. 

In this work, we investigate the effects of spectral overlapping of the Mie resonances, attributed to finite-size symmetric structures and analyze their impact on scattering enhancement. We generalize the concept of superscattering for finite-size non-spherical objects and derive a fundamental value of a single-channel limit for them. The results are supported by the full-wave numerical simulations and experimental measurements in the GHz range. In particular, we show that the SCS of a subwavelength finite-height cylindrical resonator can be substantially enhanced by the collocation of several resonances of different symmetry.

\section{Generalization of superscattering for non-spherical objects \label{sec:generalization_of_superscattering}}

T-matrix is one among semi-analytical approaches for electromagnetic scattering calculations~\cite{mishchenko1996t}. The essence of the method is the expansion of both incident and scattered fields into a series of weighted mutually orthogonal vector spherical harmonics. T-matrix links between the complex amplitudes within the expansions:
\begin{equation}
\mathbf{b}=\hat T\mathbf{a},
\label{eq:T_matrix}
\end{equation} 		     
where $\mathbf{a}$ and $\mathbf{b}$ are the vectors of complex amplitudes, corresponding to the incident and scattered fields, respectively. Each matrix element $T_{ss'}=\langle s|\hat T|s'\rangle$ is the transition amplitude between different multipole channels  (from $s$-state to  $s'$-state). The indices $s$ and $s'$ encode the polarization (electric or magnetic in respect to a chosen direction), angular momenta $\ell$ and $\ell'$ and their projections $m$ and $m'$. $\hat T$ is diagonal in the case of a spherically symmetric scattering potential. It means that each multipole of a spherical resonator is uniquely linked to a single spherical harmonic (multipole). Thus, there is no coupling between different multipoles and, for example, a dipole harmonic of the incident field can only be scattered into the dipole channel. It is worth highlighting that true eigenmodes of a structure cannot be 1:1 mapped to the scattering channels even for a spherical resonator. The reason is that different modes can have the same radiation pattern and, thus, cannot be orthogonal in the far-field. Therefore, a natural set of scattering channels is formed by vector spherical functions but not by eigenmodes of the resonator.

\begin{figure}[t]
  \centering
  \includegraphics[width=\linewidth]{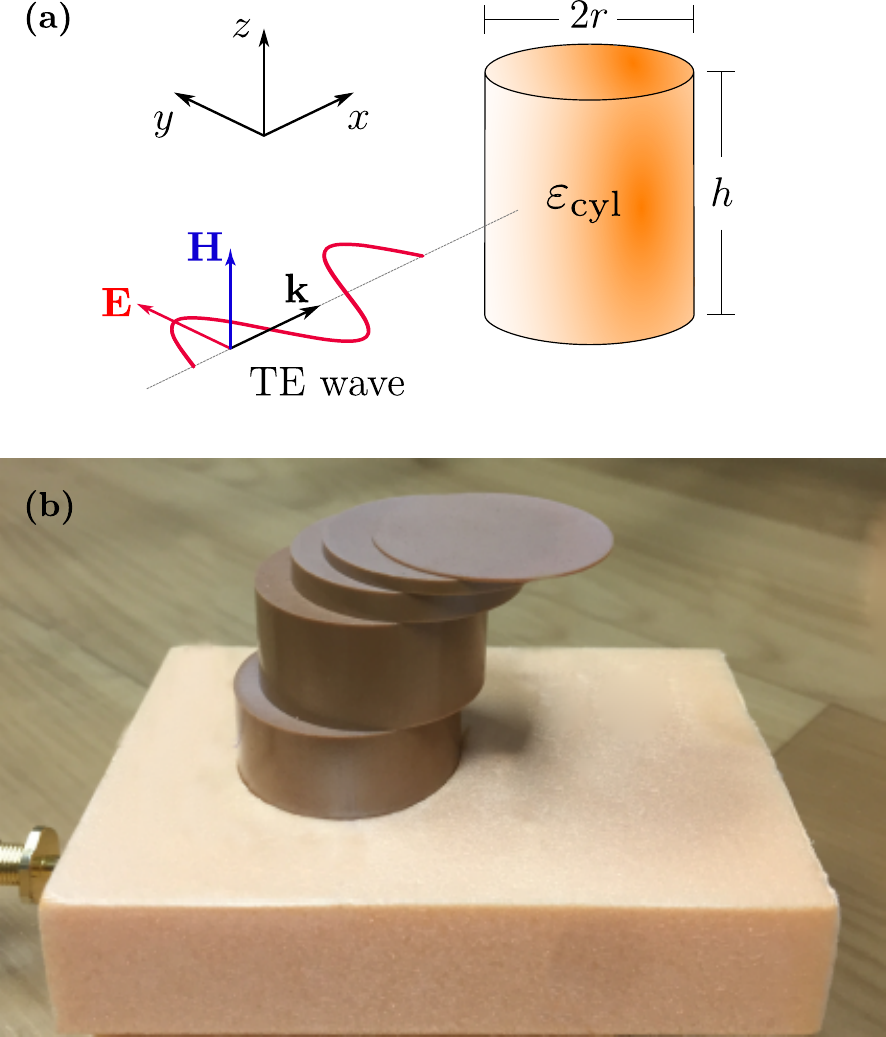}\\
  \caption{(a) Schematic geometry of the system under consideration --- a TE-polarized plane wave scattered by a cylindrical particle of height $h$ and radius $r$. (b) Experimental sample consisting of several ceramic cylinders with different heights and the same radius $15.7$~mm. Dielectric permittivity of the sample is $\varepsilon_\text{cyl} = 44.8$ and loss tangent $\tan\delta=10^{-4}$.}
  \label{fig:geometry}
\end{figure}

However, for non-spherical resonators, the multipoles contributing to the scattered field  are partially mixed, but still form independent (non-intersecting) sets. Those sets can be linked via eigenmodes of a scatterer, which can be classified with different symmetry groups. Making a re-arrangement of multipoles indices, the T-matrix of a non-spherical resonator can be presented in a block-diagonal  form~\cite{kahnert2005irreducible}: 
\begin{equation}
\hat T=\text{diag}\{\hat T_1,\hat T_2, ...\},
\label{eq:T_matrix_diag}
\end{equation} 		 
where the blocks $\hat T_s$ correspond to the modes of different symmetry. \textcolor{black}{The T-matrix of a non-spherical scatterer made of homogeneous isotropic material can be calculated rigorously using the extended boundary condition method proposed by Waterman~\cite{waterman1971symmetry}.} Two eigenmodes are affiliated with two different blocks if they are transformed differently under the operations of the resonator's symmetry group. The mixing rules and, therefore, the multipoles entering to a certain block are defined by their transformation properties. Depending on the symmetry of the resonator, the number of blocks can be either infinite or finite~\cite{schulz1999point}. If the resonator's group symmetry contains a rotation axis of infinite order (a  body of revolution, e.g. sphere, cylinder, cone, two-sphere dimer, etc), then the number of the blocks is infinite. In all other cases, this number is finite. The T-matrix, reduced to the block-diagonal, can be easily inverted. This property is extremely important for performing efficient numerical routines aiming to solve scattering problems~\cite{xiong2020constraints}.

Applying group theory approaches, one may say that the multipoles contributing to the same block of $\hat T$ form an orthogonal basis of the {\it irreducible representation} (irrep) of the resonator's symmetry group~\cite{kahnert2005irreducible}. The number of blocks is equal to the number of classes of the symmetry group or, in other words, the number of its irreducible representations~\cite{tinkham2003group}. Indeed, it is a known fact in the group theory that there is a one-to-one correspondence between the different mode types and irreducible representations (the Wigner theorem)~\cite{wigner1959group}. Therefore, the independent scattering channels of low-symmetry structures are associated with irreducible representation but not with particular multipoles as in the case of spherical resonators. Thus, the total SCS can be represented as a sum of partial SCSs corresponding to the different irreducible representations:
\begin{equation}
C_\text{sct}=\sum_{s=\text{irreps}}\sum_{\ell,m,\sigma}C_{\ell m}^{\sigma s}.
\end{equation}
Of course, this equation is also valid for spherical resonators, where the polarization $\sigma$ and angular momentum $\ell$ identify the irreducible representations of O(3) -- the group of rotations in three dimensions~\cite{varshalovich1988quantum}. The number of irreducible representations is finite for the most point symmetry groups and, thus, we have a finite number of independent scattering channels. An exception as we mentioned above is resonators with the rotation axis of infinite order. Therefore, we can conclude that in order to achieve superscattering for the non-spherical resonator, we need to provide coincidence of resonant frequencies for several modes from different irreducible representations (different blocks of T-matrix). This is possible via tuning the geometry of the resonator preserving its symmetry.
\begin{figure}[t]
  \centering
  \includegraphics[width=\linewidth]{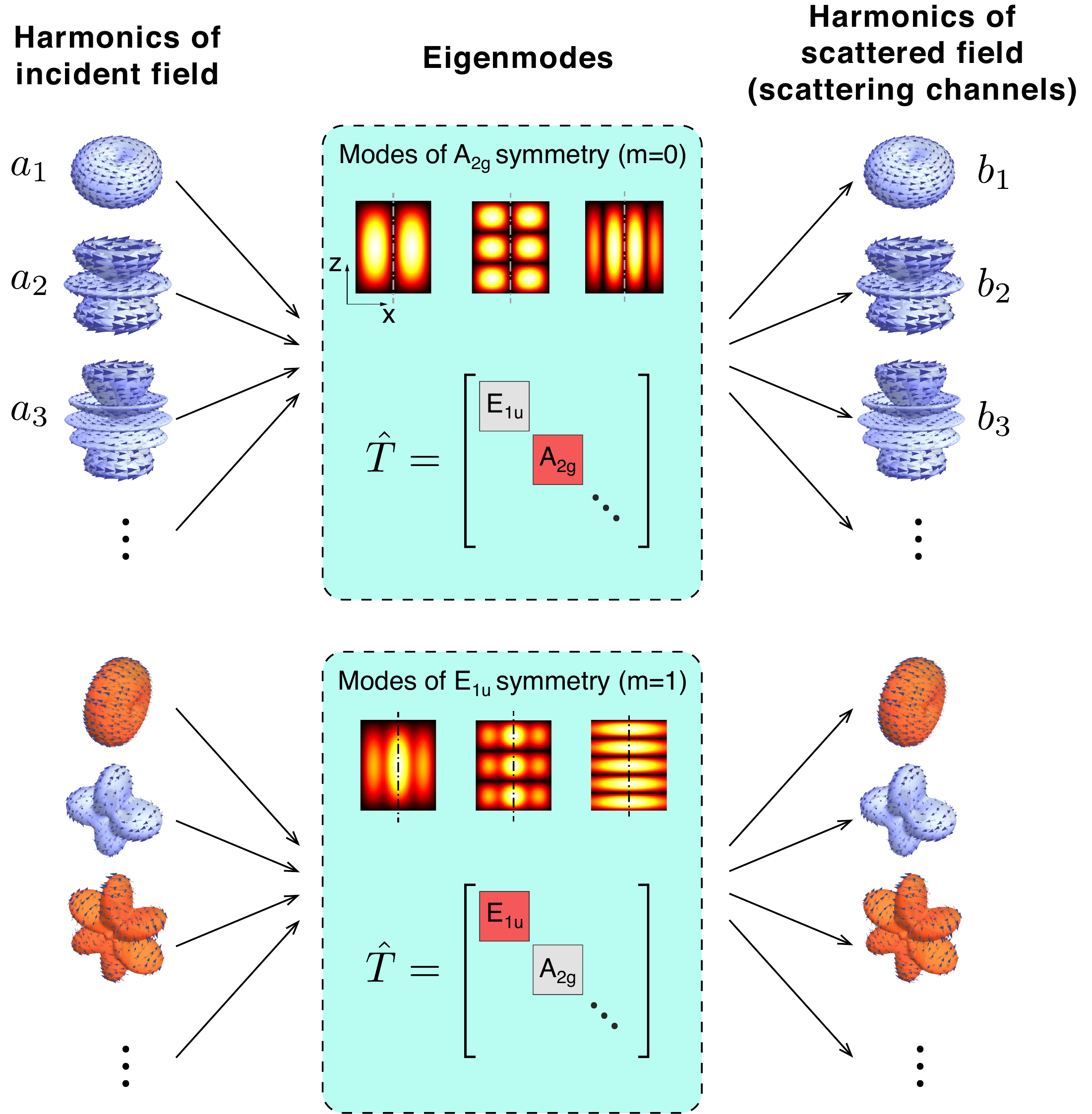}\\
  \caption{Scheme showing the relation between scattering channels corresponding to multipoles, irreducible representations of the resonator's symmetry group, and blocks of T-matrix by the example of cylindrical resonators. Here, $a_i$ and $b_i$ are the complex amplitudes of the incident and scattered fields. The insets show the characteristic mode profiles of the cylindrical resonator.}
  \label{fig:multipoles_scheme}
\end{figure}
Figure~\ref{fig:multipoles_scheme} illustrates the relation between scattering channels corresponding to multipoles, irreducible representations of the resonator's symmetry group, and blocks of T-matrix by the example of cylindrical resonator. Here, $a_i$ and $b_i$ are the complex amplitudes of the incident and scattered fields. The insets show the characteristic mode profiles of the cylindrical resonator. $\mathrm{E}_{\mathrm{1u}}$ and $\mathrm{A}_{\mathrm{2g}}$ are the standard notations of irreducible representations. Therefore, to have a superscattering for non-spherical resonator, two or more resonances corresponding to the eigenmodes from different blocks of T-matrix should be tuned to the same frequency.

\section{Maximization of scattering cross-section for cylindrical resonator}
\subsection{Numerical optimization}

\begin{figure}[htbp]
  \centering
  \includegraphics[width=\linewidth]{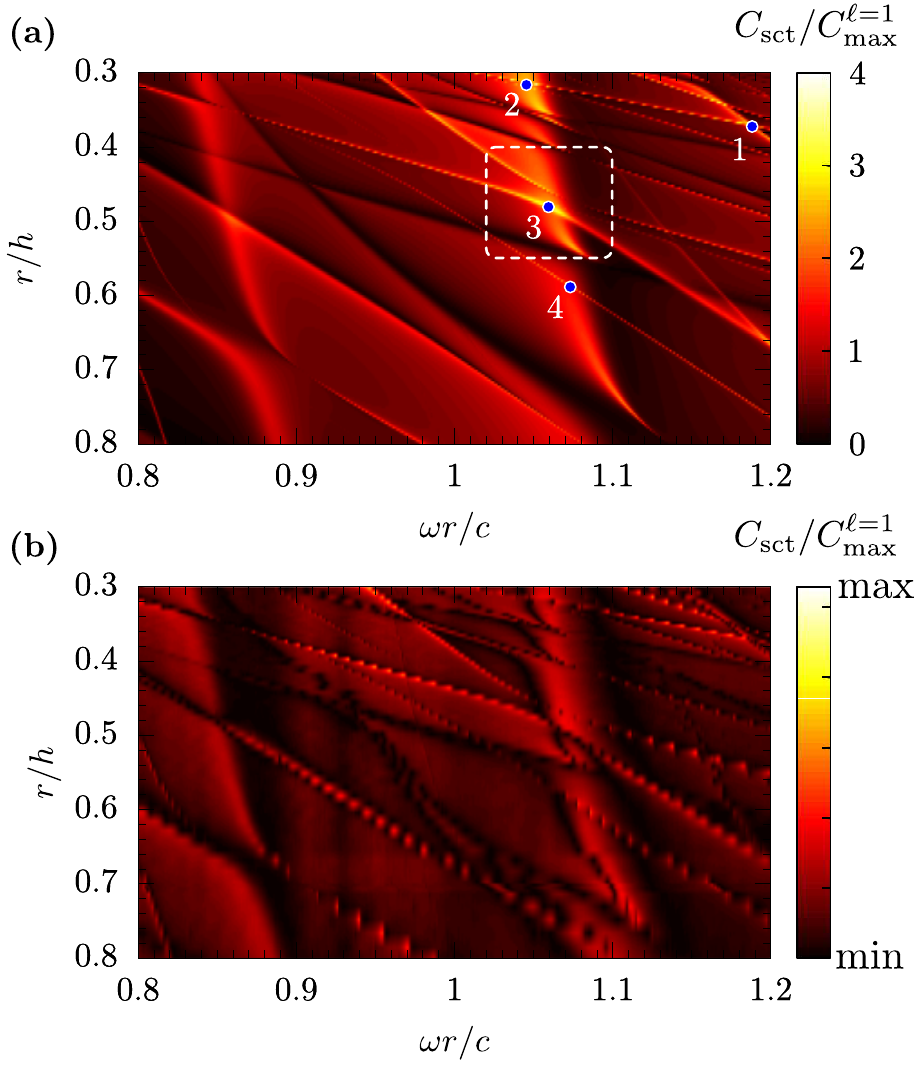}
  \caption{ Colormaps of total scattering cross-section as the function of the normalized frequency $\omega r/c$ and cylinder's aspect ratio $r/h$. Cross-sections are normalized to the corresponding dipole single-channel limit [Eq.~\eqref{eq:limit_sphere}].  Cylinder's permittivity is $\varepsilon_\text{cyl}= 44.8$ and loss tangent $\tan \delta=10^{-4}$, the incident wave is TE-polarized [see Fig.~\ref{fig:geometry}(a)]. (a) Numerical calculations. Blue dots (1-4) indicate the highest values of the normalized scattering cross-section. The white dashed rectangle indicates the area around the most pronounced maximum. (b) The measured map of the scattering cross-section spectra.}
  \label{fig:SCS_map}
\end{figure}

\textcolor{black}{To show that SCS can be enhanced by the spectral overlapping of the modes from different irreducible representations, we consider a  finite-size cylindrical resonator made of homogeneous dielectric material with permittivity $\varepsilon_{\text{cyl}}$. The resonator is illuminated by the TE-polarized incident wave, as it is shown in Fig.~\ref{fig:geometry}(a). The further results are general and applicable for a wide range of permittivities $\varepsilon_{\text{cyl}}$ and wavelengths $\lambda$ but to be specific and have an illustrative example we assume that the cylinder is made of high-refractive-index ceramics with $\varepsilon_{\text{cyl}}=44.8$ and loss tangent $\tan\delta=10^{-4}$.  Such materials demonstrate pronounced Mie resonances~\cite{bohren1998absorption,kruk2017functional,kivshar2017meta} and they are quite prospective for compact filters, antennas, wireless power transfer systems, and RFID technologies~\cite{yang2010super,sievenpiper2011experimental,peng2004dielectric,dobrykh2020long,song2016wireless,song2017wireless}.}

The calculations of the SCS spectra for different values of the aspect ratio were done with the T-matrix method and then verified with full-wave numerical simulation using the CST Microwave Studio and COMSOL Multiphysics~\cite{mishchenko1996t, bogdanov2019bound}. The numerical results are shown in Fig.~\ref{fig:SCS_map}(a), where the values of $C_\text{sct}$ are normalized to the dipolar single channel [Eq.~\eqref{eq:limit_sphere}, $\ell = 1$] to highlight the relative enhancement. The incident wave is TE polarized [electric field is perpendicular to the cylinder's axis, see Fig.~\ref{fig:geometry}(a)]. 

The color map in Fig.~\ref{fig:SCS_map}(a) demonstrates the impact of resonant modes on the SCS. The bright branches indicate the evolution of modes as the function of the system's parameters. For example, increasing the aspect ratio of the cylinder leads to reducing the resonant frequencies of its basic electric and magnetic modes. The most interesting points on the color map are those, where several modes of the cylinder are overlapped. Four typical points, marked in the plot, are chosen for a detailed analysis, and the system's parameters for those cases are summarized in Table~\ref{tab:CSC_points}. It can be seen that the SCS prevails the single-channel dipolar limit for spherical objects by a factor of 3-4 if the parameters are properly adjusted. 

\begin{table}[ht]
 \centering
  \caption{Maximal values of scattering cross-section}
  \label{tab:CSC_points}
  \begin{tabular}{@{}cccc@{}}
    \hline
    \\[-8pt]
    Point & $C_{\mathrm{sct}}/C_{\mathrm{max}}^{\ell = 1}$ & $r/h$ & $\omega r/c$ \\[3pt]
    \hline
    1  & 4.01 & 0.369 & 1.187 \\
    2  & 3.93 & 0.312 & 1.044 \\
    3  & 3.36 & 0.480 & 1.062 \\
    4  & 2.75 & 0.582 & 1.073 \\
    \hline
  \end{tabular}
\end{table}

It is important to highlight that the proposed design is very simple and practical as the scattering enhancement is achieved by tuning of only one geometrical parameter without the need to use coating layers or additional structuring of the resonator's surface as in Refs.~\cite{qian2019experimental,mirzaei2013cloaking,raad2019multi,qian2018multifrequency,mirzaei2014superscattering}. 

\subsection{Sample and experimental measurements}

Next, we provide experimental verification of the scattering enhancement in the GHz frequency range. To vary the height of the cylindrical resonator, we sliced a long ceramic rod into several sections and obtained a collection of ceramic disks with different heights and an identical diameter ($31.4$~mm).  The rod is manufactured by sintering ceramic powder of calcium titanate-lanthanum aluminate (LaAlO$_3$-CaTiO$_3$) into a solid. The permittivity of the fabricated disks is $\varepsilon_\text{cyl}=44.8$ and loss tangent $\tan{\delta}=10^{-4}$ for low GHz frequencies {\cite{nenasheva1992ceramic}}. The fabricated set of the polished disks allows changing the heights in the range from 0.25~mm to 15~mm with the step 0.2~mm. Figure~\ref{fig:geometry}(b) shows several representative examples. A dielectric foam holder has been fabricated by drilling Penoplex -- a foam material transparent to GHz waves. 

The SCSs of the structures were measured in an anechoic chamber. Wideband horn antennas (operational range 1-18~GHz) were connected to ports of Agilent E8362C VNA. Transmitting antenna excited a quasi-plane wave impinging the resonator placed in the far-field zone. The incident field was TE polarized. The scattered field in the forward direction was collected with an additional horn antenna, which was connected to the second port of the VNA. The background signal, obtained from the measurements without a sample present, was subtracted \cite{larsson2008extinction}. Additionally, the time gating technique was applied to reduce residual reflections from the horn antennas and setup elements~\cite{deporrata-doriaiyague1998analysis}.  The measured complex transmission coefficient was used to calculate the total SCS via the optical theorem exploiting the procedure described in Ref.~\cite{larsson2013wideband,odit2020observation}. Figure~\ref{fig:SCS_map}(b) shows the measured map of the SCS. The high quality of the samples and accurate experimental acquisition provide a very good agreement between the numerical and experimental data, clearly demonstrating the mode intersection at points, where the SCS is maximal.

\begin{table}[t]
  \centering
  \caption{Irreducible representations (irreps) of the symmetry group $D_{\infty h}$ and eigenmode classification. $m$ is the azimuthal number. For TM-polarized modes $\mathbf{E}=(E_\rho,0,E_z)$ and $\mathbf{H}=(0,H_\varphi,0)$. For TE-polarized modes $\mathbf{E}=(0,E_\varphi,0)$ and $\mathbf{H}=(H_\rho,0,H_z)$. For even modes $\mathbf{E}(x,y,-z)=\mathbf{E}(x,y,z)$ and for odd modes $\mathbf{E}(x,y,-z)=-\mathbf{E}(x,y,z)$.}
  \label{tab:irreps}
  \begin{tabular}{@{}ccccc@{}}
    \hline
    \\[-8pt]
    Irrep & $m$ & $\substack{\text{\small Polarization} \\ \text{ \small of mode}}$ & $\substack {\text{\small Parity:} \\ \text{\small $z\rightarrow-z$}}$ & $\substack{\text{\small Excitation} \\ \text{ \small by TE-wave}}$ \\[3pt]
    \hline
    A$_{1u}$  & 0 & TM & odd & no \\
    A$_{1g}$  & 0 & TM & even & no \\
    A$_{2u}$  & 0 & TE & odd & no \\
    A$_{2g}$  & 0 & TE & even & yes \\
    E$_{1g}$  & 1 & hybrid & odd & no \\
    E$_{1u}$  & 1 & hybrid & even & yes \\
    E$_{2u}$  & 2 & hybrid & odd & no\\
    E$_{2g}$  & 2 & hybrid & even & yes \\
    E$_{3g}$  & 3 & hybrid & odd & no \\
    E$_{3u}$  & 3 & hybrid & even & yes\\
   \vdots &  \vdots  & \vdots &  \vdots  &  \vdots  \\
    \hline
  \end{tabular}
\end{table}

\section{Modal and multipole description of scattering enhancement }

\begin{figure*}[htbp]
  \includegraphics[width=\linewidth]{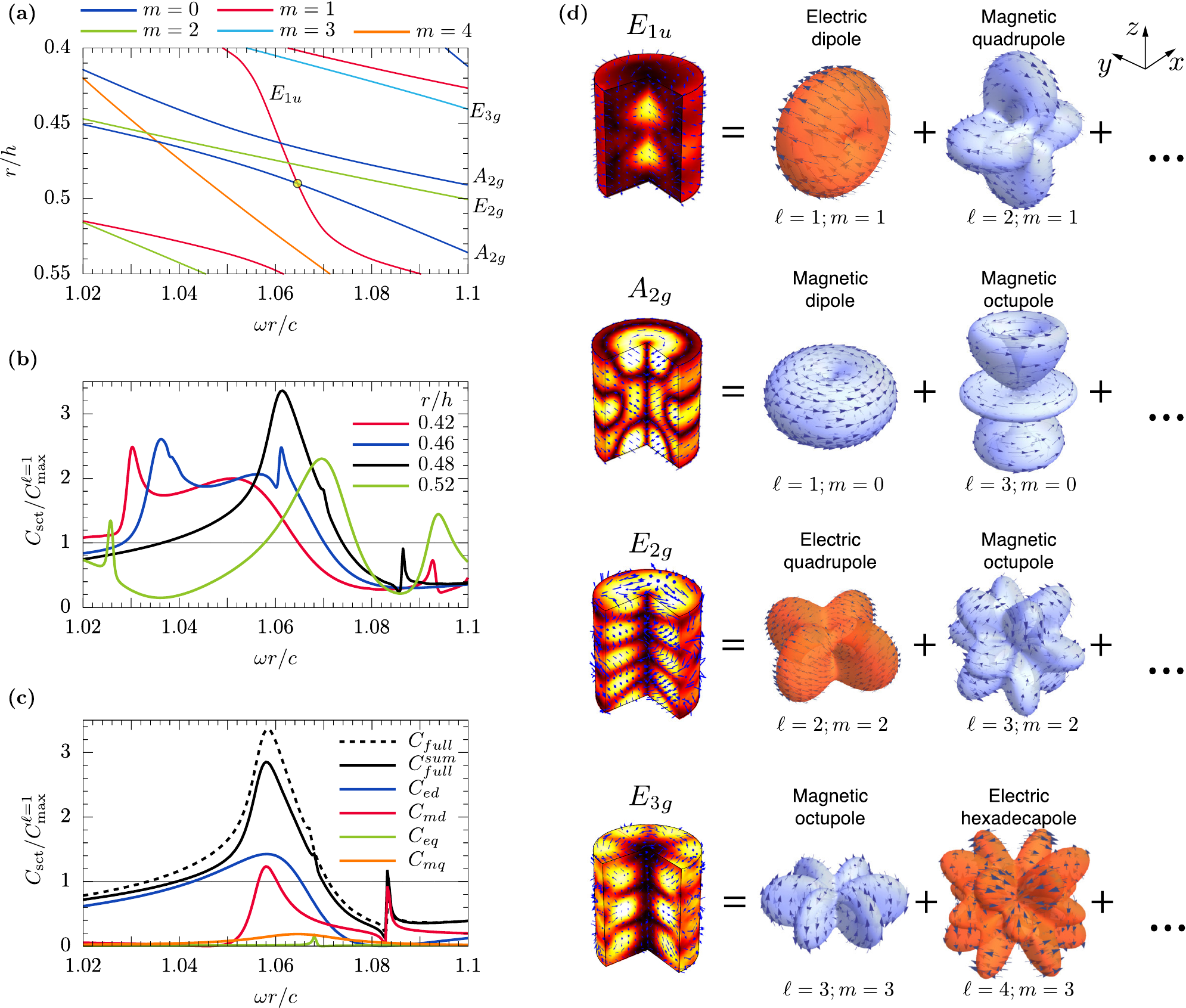}%
  \caption{(a) Spectrum of the eigenmodes calculated in the vicinity of the SCS maxima and their irreducible representations. (b) Spectrum of the SCSs, corresponding to the slices of the diagram presented in Fig.~\ref{fig:SCS_map}(a). (c) Multipole decomposition of the total SCS at $r/h = 0.48$. Black-dashed line is the result of full-wave numerical simulation, black-solid line is the result of summation of the partial SCSs (ED, MD, EQ, and MQ) shown by colored lines. (d) Examples of field distributions of eigenmodes corresponding to the considered irreducible representations, and their multipole far-field decompositions.}
  \label{fig:eigenmodes}
\end{figure*}

To show that the scattering enhancement [points 1, 2, 3, and 4 in Fig.~\ref{fig:SCS_map}(a)] is directly related to the overlapping of the eigenmodes transforming under different {\it irreducible representations}, we calculated the eigenmode spectrum as a function of aspect ratio [see Fig.~\ref{fig:eigenmodes}(a)]. These calculations were carried out using COMSOL Multiphysics software. To get only the modes that can be excited by a plane TE-polarized wave [see Fig.~\ref{fig:geometry}(a)] we perform the simulation in a quarter of space applying perfect electric boundary condition to the $xz$-plane and perfect magnetic boundary condition to the $xy$-plane. To classify the modes we analyzed their transformation rules using the character tables for point symmetry group $D_{\infty h}$ -- the symmetry group of a finite cylinder~\cite{gelessus1995multipoles, ohtaka1996photonic}. The modes of a finite cylinder can be characterized by azimuthal number $m=0,1,2,...$ and parity (odd or even) of the electric field distribution with respect to the mirror reflection $z\rightarrow-z$. For $m=0$ we can introduce the polarization of the mode (TE or TM). It is worth mentioning that an orbital number $\ell$ is no longer a good quantum number of the mode as it was in the case of spherical resonators. The classification of the modes in a finite cylinder and their connection to the irreducible representations of $D_{\infty h}$ is shown in Table~\ref{tab:irreps}.
The first column shows the notations of irreducible representations; the second column shows the azimuthal numbers -- projections of angular momentum on the $z$-axis; the third column shows the polarization of the mode (in the cases, when it can be defined);  the forth column shows the parity of the mode (odd or even) under its reflection in the $xy$-plane; the last column indicates whether it is possible to excite the mode by the TE-polarized wave propagating perpendicular to the cylinder axis. A characteristic electric field distribution and the multipole content for several modes from different irreducible representations are shown in Fig.~\ref{fig:eigenmodes}(d). The detailed mode classification and  multipole analysis in the resonators of different symmetry groups is provided in Ref.~\cite{gladishev2020symmetry}.

The modes from different irreducible representations are shown in Fig.~\ref{fig:eigenmodes}(a) with different colors. We use the standard notations of the irreducible representations~\cite{ivchenko1995superlattices}. One can see that only the modes from different irreducible representations can intersect, producing a crossing in the parametric space. The modes from the same irreducible representation repel forming an avoid crossing [see modes A$_{2g}$ in Fig.~\ref{fig:SCS_map}(a)]. In quantum mechanics, this fact is well-known as the von Neumann-Wigner theorem ~\cite{wigner1959group}. This repulsion is a common feature of open systems~\cite{wiersig2006formation} and it is explained by the interaction of the modes through the continuum of propagating waves in the surrounding space. In particular, high-Q quasi-bound states in the continuum can appear as a result of this interaction~\cite{rybin2017high, bogdanov2019bound,koshelev2020subwavelength,mylnikov2020lasing}.   

As an example, let us investigate the scattering enhancement near point 3 in Fig.~\ref{fig:SCS_map}(a) in more details. Comparing the spectral position of the maximum of SCS in Figs.~\ref{fig:SCS_map}(a) and \ref{fig:eigenmodes}(b) with the map of eigenmodes in Fig.~\ref{fig:eigenmodes}(a) one can see that the maximum in SCS appears at $r/h=0.48$ as a result of the co-location of two modes from different irreducible representations, namely, A$_{2g}$ and E$_{1u}$. The maximal value of SCS in this case prevails in a dipolar single-channel limit ($\ell = 1)$ for a spherical object by more than three times. 

Then we can consider the multipole decomposition at the point of maximal SCS, which can be done with the use of the well-known analytical expression~\cite{jackson1998classical}:
\begin{equation}
  \begin{aligned}
    C_{\mathrm{sct}} = \frac{k^4}{6\pi\varepsilon_0^2 |\mathbf{E}_0|^2} \left[\sum\limits_{\alpha} \left(|p_{\alpha}|^2 + |m_{\alpha}/c|^2 \right) + \right. & \\
    \frac{1}{120} \sum\limits_{\alpha \beta} \left(|k Q^e_{\alpha \beta}|^2 + \left|\frac{k Q^m_{\alpha \beta}}{c} \right|^2 \right) + &\\
    \left. \frac{1}{315} \sum\limits_{\alpha \beta \gamma} \left(|k^2 O^e_{\alpha \beta \gamma}|^2 + \left|\frac{k^2 O^m_{\alpha \beta \gamma}}{c} \right|^2 \right) + \ldots \right]
    \end{aligned}
    \label{eq:multipoles}
\end{equation}
where $\mathbf{E}_0$ is the amplitude of the incident plane wave, $k$ is the wavenumber, $c$ is the speed of light in free space, $\varepsilon_0$ is the vacuum permittivity, $p_{\alpha}, m_{\alpha}$ are electric and magnetic dipole moments, $Q^{e,m}_{\alpha \beta}$ are the electric and magnetic quadrupole moments, $O^{e,m}_{\alpha \beta \gamma}$ are the electric and magnetic octupole moments,  etc. The summation indices $\alpha, \beta, \gamma$ run over the Cartesian coordinates $x,y,z$. 
These multipole moments can be calculated analytically as volume integrals~\cite{alaee2018electromagnetic}. The total SCS for subwavelength objects is accurately defined by the lower multipoles. Indeed, it can be seen from Fig.~\ref{fig:eigenmodes}(c) that the dominant contribution to SCSs for the modes E$_{1u}$ and A$_{2g}$  is given by the electric and magnetic dipole moments, respectively. 

As one can see from Fig.~\ref{fig:eigenmodes}(c), the partial SCSs corresponding to the electric and magnetic dipole moments exceed the single-channel limit for a sphere [Eq.~\eqref{eq:limit_sphere}]. The reason for this is the mixing of multipoles (see Fig.~\ref{fig:multipoles_scheme}) making possible rescattering between the channels corresponding to different multipoles. Therefore, the single-channel limit should be specified for the case of non-spherical resonators. We will discuss it in Sec.~\ref{sec:single_channel_limit}.

\begin{figure}[t]
   \centering
   \includegraphics[width=\linewidth]{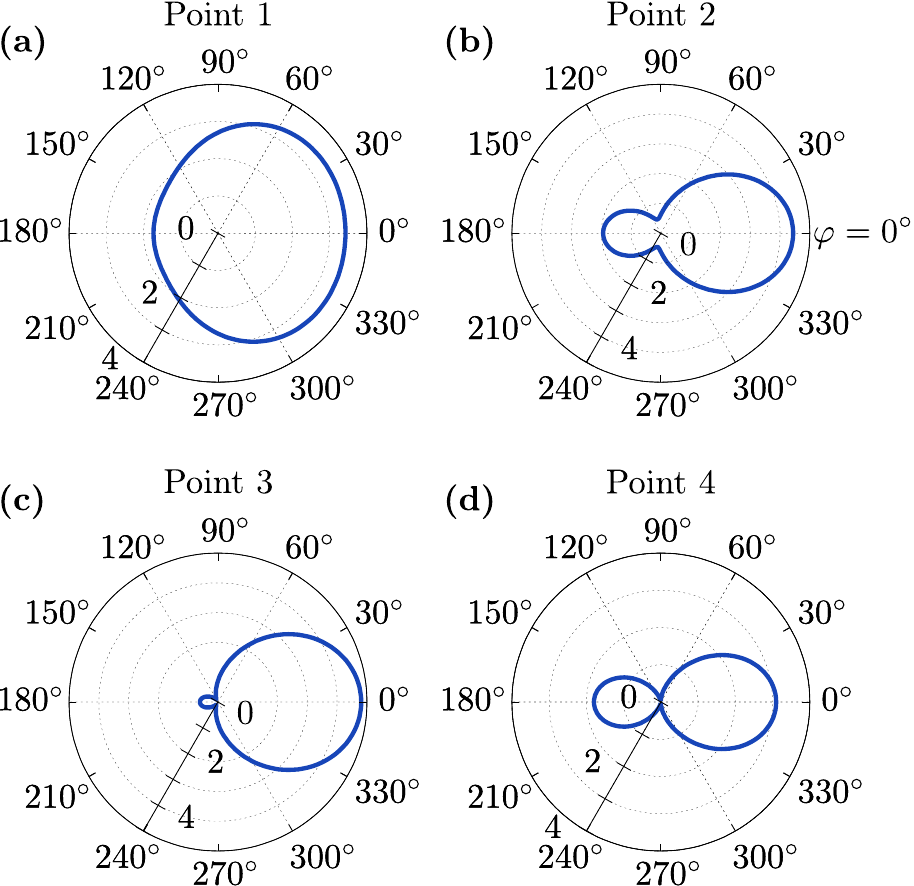}
   \caption{Directivity patterns obtained for the points corresponding to maximal values of the scattering cross-section indicated in Fig.~\ref{fig:SCS_map}(a) and listed in Table~\ref{tab:CSC_points}. The forward direction correspond to azimuthal angle $\varphi=0\degree$. The scale is linear.}
   \label{fig:directivity}
\end{figure}

According to the optical theorem, when the losses are negligible, the maximization of SCS should result in an increase of scattering in the forward direction and forward directivity. The directivity in the direction given by the polar and azimuthal angles $\theta$ and $\varphi$ is defined as~\cite{bohren1998absorption}
\begin{equation}
  D(\theta, \varphi) = \frac{4\pi I(\theta, \varphi)}{\int\limits_{0}^{2\pi}\int\limits_{0}^{\pi} I(\theta, \varphi) \sin\theta d\theta d\varphi}.
\end{equation}
Here, $I(\theta, \varphi)$ is the intensity of the scattered wave in a direction given by $\theta$ and $\varphi$. Figure~\ref{fig:directivity} shows the directivity patterns in the $xy$-plane ($\theta=\pi/2$) at the points 1, 2, 3 and 4 [see Fig.~\ref{fig:SCS_map}(a)], where the maximal total cross-section is achieved. One can see that for all points the forward directivity exceeds the value of 3 and for the point 3, the forward directivity is almost 5. This fact is quite interesting as it highlights the contribution of relative phases between multipolar contributions, which can be constructively interfere and create directive patterns~\cite{noskov2018non, barhom2019biological}.

\section{Discussion on single channel limit \label{sec:single_channel_limit}}
Single-channel limit, derived for the spherical objects [Eq.~\eqref{eq:limit_sphere}], is not valid for the particles with the symmetry of a lower order, as we have demonstrated numerically in the previous section. The underlying reason for this inconsistency can be illustrated with T-matrix, which set a relation between the vectors of the complex amplitudes of the incident and scattered waves usually written in the basis of vector spherical harmonics  [see Eq.~\eqref{eq:T_matrix}]. 

In the general case, T-matrix can be written in a block diagonal form [see Eq.~\eqref{eq:T_matrix}], such that each block $T_s$ corresponds to the modes of certain symmetry, i.e. from a certain irreducible representation, and the elements of the block are the transition amplitudes between different scattering channels. For example, a quadrupole harmonic from the incident plane wave can be scattered into the dipole channel. This is in a sharp contrast to the case of spherical resonators, where the T-matrix is diagonal and, therefore, there is no rescattering between different channels. This multipole mixing inspired by non-spherical potentials is the reason why the scattering into a single channel can surpass the limit of the same channel for a spherical object. 

To generalize the single-channel limit for non-spherical objects we rely on singular value decomposition~\cite{eckart1939principal} and the fact that the SCS can not exceed the extinction cross-section. Following the derivation procedure presented in Appendix A, an upper limit of partial SCS corresponding to angular momentum $\ell$ can be determined as 
\begin{equation}
  C^{\ell}_{\mathrm{max}} = \frac{ 2\ell + 1}{\pi} \lambda^2,
\end{equation}
which is only twice bigger than the single-channel limit for spherical particles. We should emphasize that the question of whether this upper bound can be reached, and, if yes, then under what conditions, remains open and it will be the subject of further research. Another important point is that the blocks of T-matrix have an infinite size and, therefore, the modes contribute to the infinite number of scattering channels. Nevertheless, the SCS remains finite due to the contribution of high-order multipoles are negligible for the finite-size objects. \textcolor{black}{Indeed, in practice, the field expansions and the T-matrix are terminated at some angular momentum $\ell_{\text{max}}$. If a scatterer is contained within a radius $a$, then the number $\ell_{\text{max}}\approx \omega a/c$~\cite{brock2000using, opsal1985theory}.} 

\textcolor{black}{The T-matrix approach formulated in terms of the vector spherical harmonics is not a unique method for solving the scattering problem. At first sight, it may seem that a more natural way is to take a complete set of the eigenfunctions of a resonator as a basis. However, it is well-known that for open systems, their eigenfunctions (resonant states) diverge at infinity~\cite{doost2014resonant, lalanne2018light}. They are not normalized in a regular way that makes them less convenient in practice. Nevertheless, having a spectrum of the system and the complete set of resonant states one can rigorously derive the expression for the scattering matrix of the system or for the total scattering cross-section~\cite{alpeggiani2017quasinormal,weiss2018calculate,koshelev2017theory}.}

\textcolor{black}{Resonant states can not be associated with the scattering channels because, for example, two different modes can have  completely the same far-field angular distributions. Therefore, their far-fields are not orthogonal and interfere, and the total scattered power by these modes cannot be divided into two independent terms. In contrast, using a basis of vector spherical harmonics, the total scattering cross-section can be represented as a sum of partial cross-sections corresponding to different polarization, orbital momentum, and its projection. Thus, such an approach is more practical and it gives a deeper physical insight into the scattering problem without direct appeal to the eigenmodes. The question of how much energy can be scattered in by a single mode is also open to the best of our knowledge.}

\section{Conclusion}
We have generalized the concept of superscattering for non-spherical resonators. The main strategy for scattering enhancement is spectral overlapping between eigenmodes of different symmetry. This can be achieved by tuning geometrical parameters of an object, while still preserving its symmetry. The scattering enhancement due to spectral overlap of was demonstrated by the example of a high-index dielectric resonator of a finite height and it was shown to prevail a dipolar single-channel for a spherical scatterer up to a factor of four.  The tuning parameter was the cylinder’s aspect ratio and no coating layers or additional structuring were used. This approach considerably simplifies the design, making it attractive in a broad range of possible applications. The obtained results allow designing compact superscatterers and keep in mind fundamental limitations, which might be faced during the maximization of the performances.  We have also demonstrated that the partial SCS of a non-spherical object, corresponding to a certain multipole, can exceed the single-channel limit for a sphere for the same multipole. Using the singular value decomposition and the fact that the SCS can not exceed the extinction cross-section we have shown the single-channel limit for non-spherical objects is twice bigger than one for spherical resonators. However, the specific conditions for reaching this limit remains open. 

\acknowledgements
The Authors thank Kristina Frizyuk for fruitful discussions and Elizaveta Nenasheva (CEO of Ceramics Co., Ltd.) for sample fabrication. The research was supported by the Russian Science Foundation (Project 19-79-10232). A.B. acknowledges the RFBR (19-02-00419) and the BASIS foundation.

\appendix 
\section{Single-channel limit for non-spherical scatterers}

In this appendix we derive a general upper limit on a scattering channel
of the scattering cross-section (SCS). The T-matrix by definition relates
amplitude vectors $\bm{a}$ and $\bm{b}$ of incident (incoming) and
scattered (outgoing) fields respectively:
\begin{equation}
\bm{b}=\hat T\bm{a}
\end{equation}
The fact that the scattering power is less than the extinction power
is known to lead to the relation (e.g., see \cite{mishchenko2000tmat}, Chapter 6)
\begin{equation}
\bm{a}^{\dagger}\hat T^{\dagger}\hat T\bm{a}\leq-\dfrac{1}{2}\bm{a}^{\dagger}\left(\hat T+\hat T^{\dagger}\right)\bm{a}
\label{eq:tmat_ineq}
\end{equation}
for any incident field vector $\bm{a}$, where the dagger means Hermitian conjugation of a matrix. The equality holds for nonabsorbing particles.

Let us consider the singular value decomposition (SVD) of the T-matrix $\hat T = \hat U\hat\Sigma \hat{V}^{\dagger}$
with unitary matrices $\hat U^{\dagger}$ and $\hat V^{\dagger}$ (each being
composed of a set of orthonormal basis column vectors, $\hat V=\left[\bm{v}_{1}\thinspace\bm{v}_{2}\thinspace\bm{v}_{3}\thinspace\dots\right]$,
$\hat U=\left[\bm{u}_{1}\thinspace\bm{u}_{2}\thinspace\bm{u}_{3}\thinspace\dots\right]$),
and diagonal matrix $\hat\Sigma$ of singular values $\sigma_{k}$.
Since $\bm{a}$ can be any let us take $\bm{a}=\bm{v}_{k}$ in Eq. (\ref{eq:tmat_ineq}):
\begin{equation}
\bm{v}_{k}^{\dagger}\hat V\hat\Sigma^{\dagger}\hat\Sigma \hat V^{\dagger}\bm{v}_{k}\leq-\dfrac{1}{2}\bm{v}_{k}^{\dagger}\left(\hat U\hat\Sigma \hat V^{\dagger}+\hat V\hat\Sigma^{\dagger}\hat U^{\dagger}\right)\bm{v}_{k}.
\end{equation}
The product $\hat{V}^{\dagger}\bm{v}_{k}= {\bm{e}}_{k}$ is the $k$-th column of the unit matrix $\hat{I} = [{\bm{e}}_{1}\thinspace{\bm{e}}_{2}\thinspace{\bm{e}}_{3}\thinspace\dots]$. Then,
\begin{equation}
\left|\sigma_{k}\right|^{2}\leq-\dfrac{1}{2}\left(\bm{v}_{k}^{\dagger}\hat U\hat\Sigma{\bm{e}}_{k}+{\bm{e}}_{k}^{\dagger}\hat\Sigma^{\dagger}\hat U^{\dagger}\bm{v}_{k}\right)
\end{equation}
Denote $\hat W=\hat U^{\dagger}\hat V$, which has the unitary property $\hat W^{\dagger}\hat W=\hat I$,
with $k$-th column $\bm{w}_{k}=\hat U^{\dagger}\bm{v}_{k}$:
\begin{equation}
\begin{split}
\left|\sigma_{k}\right|^{2}&\leq-\dfrac{1}{2}\left(\sigma_{k}\bm{w}_{k}^{\dagger}{\bm{e}}_{k}+\sigma_{k}^{*}{\bm{e}}_{k}^{\dagger}\bm{w}_{k}\right)\\ &= -\text{Re}\left\{ \sigma_{k}\left(\bm{w}_{k}^{\dagger}{\bm{e}}_{k}\right)\right\}
\end{split}
\label{eq:sigma_less1}
\end{equation}
For any two complex numbers $z_{1,2}=x_{1,2}+iy_{1,2}$
\begin{equation}
\left|\text{Re}\left(z_{1}z_{2}\right)\right|=\left|x_{1}x_{2}-y_{1}y_{2}\right|\leq\left|x_{1}\right|\left|x_{2}\right|+\left|y_{1}\right|\left|y_{2}\right|
\end{equation}
Since $\bm{w}_{k}^{\dagger}{\bm{e}}_{k}$ is a projection of $\bm{w}_{k}$, having a unit norm, onto the ${\bm{e}}_{k}$, then $\left|\bm{w}_{k}^{\dagger}{\bm{e}}_{k}\right|\leq1$,
which means $\left|x_{2}\right|\leq1$, $\left|y_{2}\right|\leq1$
in the latter equation. Thus, the inequality of Eq. (\ref{eq:sigma_less1}) becomes
\begin{equation}
\left|\sigma_{k}\right|^{2}\leq\left|\text{Re}\left\{ \sigma_{k}\left(\bm{w}_{k}^{\dagger}{\bm{e}}_{k}\right)\right\} \right|\leq\left(\left|\text{Re}\,\sigma_{k}\right|+\left|\text{Im}\,\sigma_{k}\right|\right)
\label{eq:ineq}
\end{equation}
The latter inequality can be shown to have solutions only when $\left|\sigma_{k}\right|^{2}\leq2$. Note, that in case of a spherically symmetric scatterer, the T-matrix is diagonal. Hence, $\hat{V}$, $\hat{U}$, and $\hat{W}$ are identity matrices, and inequality of Eq. (\ref{eq:ineq}) reduces to $|\sigma_k|^2\leq -\text{Re}\,\sigma_k$ with the solution $|\sigma_k|^2\leq1$, which was previously found for the case of Mie scattering on the basis of an analysis of Mie coefficients.

The total averaged scattering power being written via the T-matrix
SVD becomes
\begin{multline}
W_\text{sct}=\dfrac{\bm{a}^{\dagger}\hat T^{\dagger}\hat T\bm{a}}{2k^{2}Z}=\dfrac{\bm{a}^{\dagger}\hat V\hat\Sigma^{\dagger}\hat\Sigma \hat V^{\dagger}\bm{a}}{2k^{2}Z}=\dfrac{\tilde{\bm{a}}^{\dagger}\hat\Sigma^{\dagger}\hat\Sigma\tilde{\bm{a}}}{2k^{2}Z}\\
=\dfrac{1}{2k^{2}Z}\sum_{L,\sigma}\left|\sigma_{L,\sigma}\right|^{2}\left|\tilde{a}_{L,\sigma}\right|^{2}\leq\dfrac{1}{k^{2}Z}\sum_{L,\sigma}\left|a_{L,\sigma}\right|^{2}
\end{multline}
where $L=\left(\ell,m\right)$ is the spherical harmonic index, $Z=\sqrt{\mu_{s}/\varepsilon_{s}}$ is the  wave impedance of the homogeneous isotropic lossless surrounding medium, $\sigma=e,h$ encodes the
polarization of the vector spherical harmonics, and $\tilde{\bm{a}}=\hat V^{\dagger}\bm{a}$,
which means that $\left|\tilde{\bm{a}}\right|^{2}\leq\left|\bm{a}\right|^{2}$
since $\hat V^{\dagger}\bm{a}$ is the orthogonal subspace projection.
Therefore, each partial term in the $W_\text{sct}$ is bounded by
\begin{equation}
W_{\text{sct},L,\sigma}\leq\dfrac{1}{k^{2}Z}\left|a_{L,\sigma}\right|^{2}.
\end{equation}

In case of the plane wave incidence along the $z$-axis with a fixed polarization
and the electric field amplitude $E_{0}$
\begin{equation}
\left|a_{\ell,m,\sigma}\right|=\delta_{m,\pm1}\sqrt{\pi\left(2\ell+1\right)}E_{0},
\end{equation}
which yields the partial contribution to the scattering power
\begin{equation}
W_{\text{sct},\ell,\sigma}\leq\dfrac{1}{k^{2}Z}\sum_{m=\pm1}\left|a_{\ell,m,\sigma}\right|^{2}=\dfrac{2\pi\left(2\ell+1\right)E_{0}^{2}}{k^{2}Z}.
\end{equation}
The corresponding partial contribution to the SCS
is
\begin{equation}
C_{\text{sct},\ell,\sigma}=\dfrac{W_{\text{sct},\ell,\sigma}}{\frac{1}{2Z}E_{0}^{2}}\leq\dfrac{\left(2\ell+1\right)}{\pi}\lambda^{2}.
\end{equation}
This limit is twice higher than the one for spherical objects.

\bibliography{references.bib}

\end{document}